\documentclass[letterpaper, 10 pt, conference]{ieeeconf}  

\IEEEoverridecommandlockouts                              

\usepackage{graphics} 
\usepackage{epsfig} 

\title{\LARGE \bf
Multivariate Empirical Mode Decomposition of EEG for Mental State Detection at Localized Brain Lobes
}

\author{Monira Islam, Tan Lee
\thanks{Department of Electronic Engineering, The Chinese University of Hong Kong, Sha Tin, New territory, Hong Kong
        {\tt\small (email:1155131474@link.cuhk.edu.hk, tanlee@ee.cuhk.edu.hk)}}%
}

\begin{document}

\maketitle
\thispagestyle{empty}
\pagestyle{empty}

\begin{abstract}

In this study, the Multivariate Empirical Mode Decomposition (MEMD) approach is applied to extract features from multi-channel EEG signals for mental state classification. MEMD is a data-adaptive analysis approach which is suitable particularly for multi-dimensional non-linear signals like EEG. Applying MEMD results in a set of oscillatory modes called intrinsic mode functions (IMFs). As the decomposition process is data-dependent, the IMFs vary in accordance with signal variation caused by functional brain activity. Among the extracted IMFs, it is found that those corresponding to high-oscillation modes are most useful for detecting different mental states. Non-linear features are computed from the IMFs that contribute most to mental state detection. These MEMD features show a significant performance gain over the conventional tempo-spectral features obtained by Fourier transform and Wavelet transform. The dominance of specific brain region is observed by analysing the MEMD features extracted from associated EEG channels. The frontal region is found to be most significant with a classification accuracy of $98.06\%$. This multi-dimensional decomposition approach upholds joint channel properties and produces most discriminative features for EEG based mental state detection.
\newline

\end{abstract}

\section{INTRODUCTION}

Mental state can be assessed by observing a person's behaviour, mood, perception in the context of reality and detected from functional brain activities. Electroencephalography (EEG) has drawn much attention in mental state analysis because of their instant responses to cognitive activity. EEG based functional brain connectivity analysis estimates useful tool to find relation between brain activity and mental state changes \cite{c1}. 
The accessible low-cost signal acquisition process and natural response to mental activity make EEG a preferred medium for automatic analysis and detection of mental state. The present study tackles the problem of classification of two prominent mental states, namely the relax state and the working sate, based on multi-channel EEG. In this work, relax state refers to the calmness of mind whereas alertness or attention of mind to solve problem is defined as working state.  
The cerebral cortex of brain is composed of the frontal lobe, temporal lobe, parietal lobe and occipital lobe.   
In this study, the brain regions that actively responds to mental state variation are to be identified.

Conventionally temporal and spectral features were extracted from EEG signals of individual channels for mental state estimation \cite{c2,c3,c4,c5}. In \cite{c2}, a set of statistical and entropy-based features were derived from multi-channel EEG. EEG can be characterized with the rhythmic activity of the oscillatory bands, i.e. delta, theta, alpha, beta and gamma bands. In \cite{c3}, inter-dependence between EEG band-power and mental state was investigated in which spectral features of high oscillating bands were found representative. The study in \cite{c4, c5} utilized the representative spectral features to classify mental states. Fahimi et al. proposed a deep learning approach using EEG band-power features \cite{c6}. A 2D CNN followed by LSTM was adopted in \cite{c7} and an end-to-end 3D CNN model was utilized with temporal features to perform EEG based mental state classification in \cite{c8}. For this particular task, performance of the classifiers depends on the appropriate selection of dominant features \cite{c9}. 

EEG signal generation is believed to be a highly non-linear and non-stationary process \cite{c10}. Previous work suggested that, the assumption of short-time stationarity and use of pre-selected linear basis function for spectral analysis is not appropriate for EEG. Empirical Mode Decomposition (EMD) was applied on individual channels independently to derive non-linear features from IMFs for emotion analysis in \cite{c11}. In \cite{c12}, it was shown that the inclusion of
spatial information in multi-channel EEG could help to improve the performance. This motivates
our current investigation on using the multivariate EMD (MEMD) to exploit across-channel information \cite{c13}. The
MEMD determines common oscillatory modes of EEG channels preserving joint channel properties with same number of scale-aligned IMFs. The local non-linear discriminative features are obtained from those IMFs for detecting mental states. The MEMD based non-linear EEG features are processed by an ensemble classifier model which comprises Random Forest and AdaBoost. The hybrid MEMD-DWT variational features are also experimented in view of improving the classification performance. 

The rest of the paper is organised as follows. Section 2 explains MEMD in analyzing multidimensional signal. Feature extraction and classification are discussed in Section 3. Section 4 analyses experimental results. Conclusions are stated in Section 5.

\section{MULTIVARIATE EMPIRICAL MODE DECOMPOSITION}


EMD is an iterative data-dependent signal processing technique proposed by Huang et al. \cite{c13}. It decomposes a signal into a set of intrinsic mode functions (IMFs) that represent the oscillatory modes embedded in the signal. The decomposition method requires extraction of local extremes and local mean of the signal. 
EMD on multi-channel EEG hindered by non-uniformity in the number of IMFs for each channel and scale alignment problem across data channels. In MEMD, the obtained IMFs are expected to retain same rotational modes and uniformity in the frequency scale that is essential to analyze multidimensional signal. 
The steps of MEMD algorithm are as follows \cite{c13}.

Step 1: Given an $n$-variate signal $x(t)$, determine $V$ direction vectors that uniformly sample the $n$-sphere. Let the direction vectors be denoted as $s_{\theta _v}, v=1,...,V$;

Step 2: Project $x(t)$ along each of the direction vectors. The projected signals are denoted as $\left \{ p_{\theta _v}(t) \right \}_{v=1}^{V}$;

Step 3: For each of the projected signals, locate the maxima denoted as $\left \{ t_{\theta _v}^{i} \right \}_{v=1}^{V}$;

Step 4: Interpolate $[t_{\theta _v}^{i}, x\left (t_{\theta _v}^{i} \right )]$ via cubic splines to obtain multivariate envelopes $\left \{ e_{\theta _v}(t) \right \}_{v=1}^{V}$;

Step 5: Compute the local mean of the multidimensional envelopes,
$$
m(t)=\frac{1}{V}\sum_{v=1}^{V}e_{\theta _v}(t)\eqno{(1)};
$$

Step 6: Extract the signal detail as $d(t)=x(t)-m(t)$. If $d(t)$ fulfills the stoppage criterion of multivariate IMF then $d(t)=IMF$, otherwise Let $x(t)=d(t)$ and the whole process is repeated to obtain the detail.

Step 7: Subtract $d(t)$ from $x(t)$ as $x(t):= x(t)-d(t)$, where $d(t)=IMF$ and return to step 1 to continue the sifting process. Stop shifting when no more extremes can be found after a certain number of iterations and return a monotonic function.

Overall, the MEMD decomposition of signal $x(t)$ can be expressed as

$$
x(t)=\sum_{j=1}^{M}c_j(t)+r(t)\eqno{(2)}
$$

where, the n-variate IMFs, $\left \{ c_j(t) \right \}_{j=1}^{M}$ contain scale-aligned joint rotational modes and $r(t)$ is the residue.

\section{FEATURE EXTRACTION AND CLASSIFICATION}

\subsection{IMF Selection and Feature Extraction}

IMF selection is essential as some of the oscillatory modes may provide less relevant information. It is assumed that, an IMF closely related to the raw EEG is generally important for performing different tasks. The most relevant IMFs are determined from Pearson correlation coefficient between each individual IMF and the raw EEG. The importance of an IMF is measured according to their achieved individual performance of mental state classification. The following features are extracted from IMFs. They are referred to as non-linear MEMD features.

The Hjorth parameter is given as the normalized slope descriptor that characterizes the activity, mobility and complexity in EEG \cite{c14}. The coefficient of variation $(V_r)$ depicts the amplitude variation of the signal in terms of deviation from the mean value. The fluctuation index quantifies the change of intensity between two consecutive IMFs, e.g., IMF1 and IMF2 \cite{c11}. The fractal dimension (FD) parameter specifies the complexity and self-similarity of a signal and can be measured using the Higuchi algorithm \cite{c15}. Skewness and Kurtosis are statistical parameters that measure the degree of asymmetry or peakedness of data distribution. The Shannon entropy and log-energy entropy are used to measure how much information is being carried by a signal \cite{c2}. Details of these features are explained in \cite{c2,c11,c14,c15}. The MEMD features extracted from the IMFs are ranked based on their individual performance in the classification task. The combination of highly ranked features, i.e., attaining high classification accuracy, are taken as the selected MEMD features for mental state detection.
 
The effectiveness of MEMD features is evaluated in comparison to the DWT or DFT features. The {\bf Daubechies 4 (db4)} wavelet basis function is used for wavelet decomposition to decompose EEG into four subbands. A set of features, i.e. maximum, minimum, kurtosis, skewness, variance and energy, are derived from the wavelet coefficients of different bands. They are termed as the DWT features. Furthermore, the bandpower is obtained from delta, theta, alpha, beta and gamma band signal. The bandpower of these five frequency band signal and spectral entropy are regarded as the DFT features in this analysis.
 
\subsection{Ensemble Classifier}

Random Forest and AdaBoost are two widely used ensemble classifier models. In this study, these two classifiers are combined. Adaboost is a successor of gradient boosting algorithm that takes Random Forest (RF) as the base estimator to generate better prediction model. RF is a parallel tree-growing technique that learns from randomized sub-samples and estimate the outcome based on majority voting from decision trees at each iteration. AdaBoost focuses on training upon mis-classified samples. Starting from uniform weights, it sequentially adds new base estimator and re-distribute the training samples. After each iteration the weights are reversed or adjusted until the training samples are correctly classified. The accuracy is obtained from majority voting of the base classifiers weighted by their individual accuracy. Details of the training algorithm are explained below.

Let $m$ denote the number of training samples. The initial weights of the samples are denoted as $W_1 (i)=\frac{1}{m}$. Weak hypothesis of the classifier is $h_t(x):x\to \left\{\pm 1 \right\}$ for t=1, 2,...,T which use the distribution $W_t$. The weights on training samples are updated as,

$$
W_{t+1}(i)=\frac{W_texp(-\alpha_ty_ih_t(x_i) )}{Z_t}\eqno{(3)}
$$
where $Z_t$ is a normalization factor.  $x_i$ and $y_i$ are training samples and output labels respectively. $\alpha_t$  is the weight of the classifier and it is computed from low weighted error rate.

$$
\alpha_t=\frac{1}{2}ln\left ( \frac{1-\xi _t}{\xi _t} \right )\eqno{(4)}
$$
Output hypothesis of the final estimator can be expressed as,

$$
H(x)=sign\left ( \sum_{t=1}^{T}\alpha _th_t(x) \right )\eqno{(5)}
$$

\section{EXPERIMENTS AND RESULTS}

\subsection{EEG Database and Experimental Setup}

A publicly available EEG database is used for mental status detection. The database is accessible via the Mendeley Data repository \footnote{https://data.mendeley.com/datasets/8c26dn6c7w/1}. The database consists of EEG recordings from 30 healthy subjects (age: 18-20 yrs; gender: 57\% male and 43\% female). Each subject participated in 4 sessions of recording, from each of which a continuous 3-minute EEG signal was obtained. In each recording, the first minute and the third minute are for the relax state. A task of visual problem solving was performed during the second minute, which stimulates the working state. Therefore there are a total 120 three-minute recording sessions. 
For data acquisition, a 14-channel EMOTIV EPOCH+ headset \footnote{https://emotiv.gitbook.io/epoc-user-manual/} was used where AF3, AF4 F7, F8, F3 and F4 are $6$ frontal channels. FC5, T7, T8, FC6 are $4$ temporal ones whereas P7, O1, O2, P8 channels are considered as parietal. The sampling frequency was 128 Hz and a filter with pass-band of $2$-$45$ Hz was used to suppress the high-frequency artifacts and power line noise. All EEG signals were segmented into short-time frames of $15$ second long with $10$ second overlap. Subject-independent analysis was carried out with repeated 5-Fold cross validation where 80\% of total samples on trail level were considered for training and 20\% for testing. Final result would be average of the accuracy of each fold.

\subsection{Performance of MEMD Features}
\begin{figure}[thpb]
  \centering
  {\includegraphics[width=8.0cm, height=4.6cm]{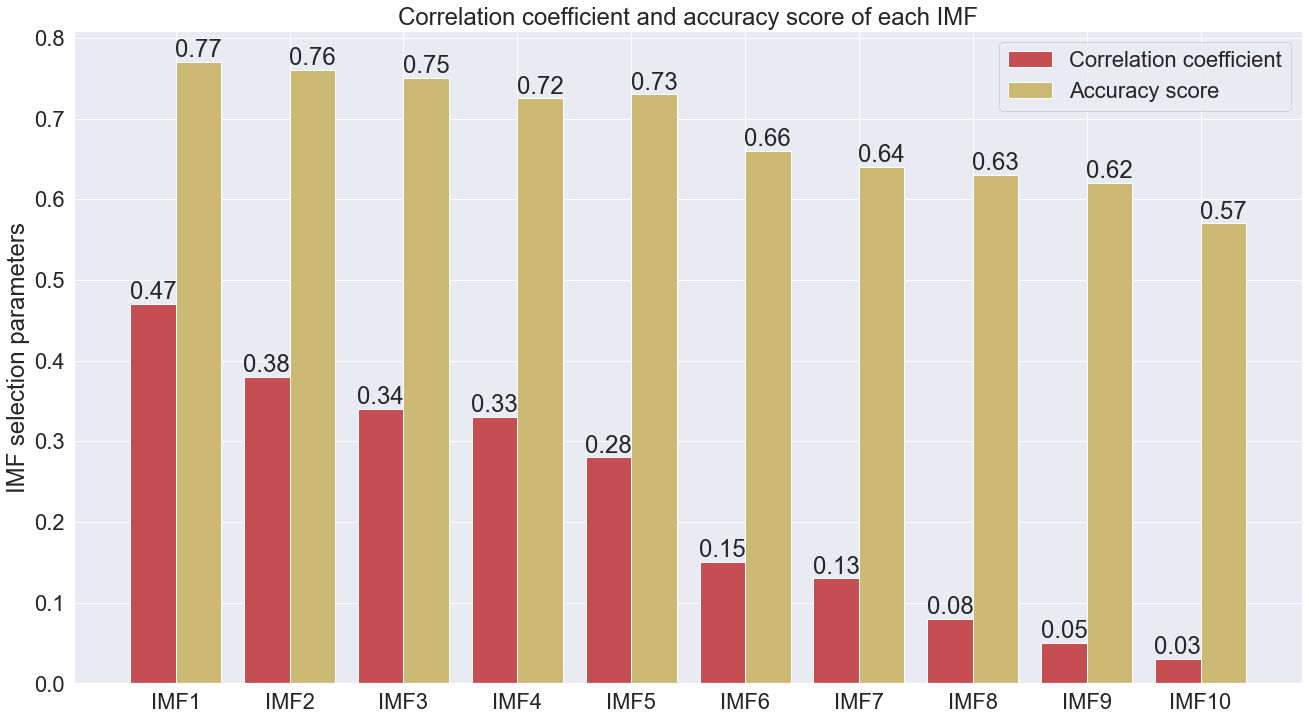}}
  \caption{Correlation coefficients and classification accuracy attained with individual IMFs}
    \label{fig:res2}
\end{figure}

For emotion detection, non-linear MEMD features are derived from a few selected IMFs. MEMD is applied to decompose EEG signals into $10$ IMFs. The lower-order IMFs represent high-oscillation components and the higher-order IMFs are low-oscillation ones. The dominance of individual IMFs is determined by the respective performance attained on mental state classification. The correlation between each IMF and the raw signal is considered. The relevant IMFs are highly correlated with the original signal and yield useful features for detecting the mental states. Using Pearson correlation formulae the correlation coefficients between each IMFs and original EEG signal is obtained as illustrated in Fig. ~\ref{fig:res2}. It is observed that the IMFs with strong correlation can deliver effective features to detect mental states. Fig. ~\ref{fig:res2} shows that the first decomposed component, i.e., IMF1 is highly correlated to the raw EEG and achieves an accuracy of $77\%$ on its own.  The classification performance degrades drastically for higher-order IMFs. IMF10 alone gives a low accuracy of $57\%$ only. The features derived from IMFs $6-10$ are found non-discriminatory for this task. This suggests that high-oscillation IMFs are most useful in reflecting the mental states. Based on the correlation coefficient and individual performance, the combination of the first five IMFs, i.e., IMF1 to IMF5, are selected for the intended task of mental state classification.

\begin{figure}[thpb]
  \centering
  {\includegraphics[width=8.0cm, height=4.5cm]{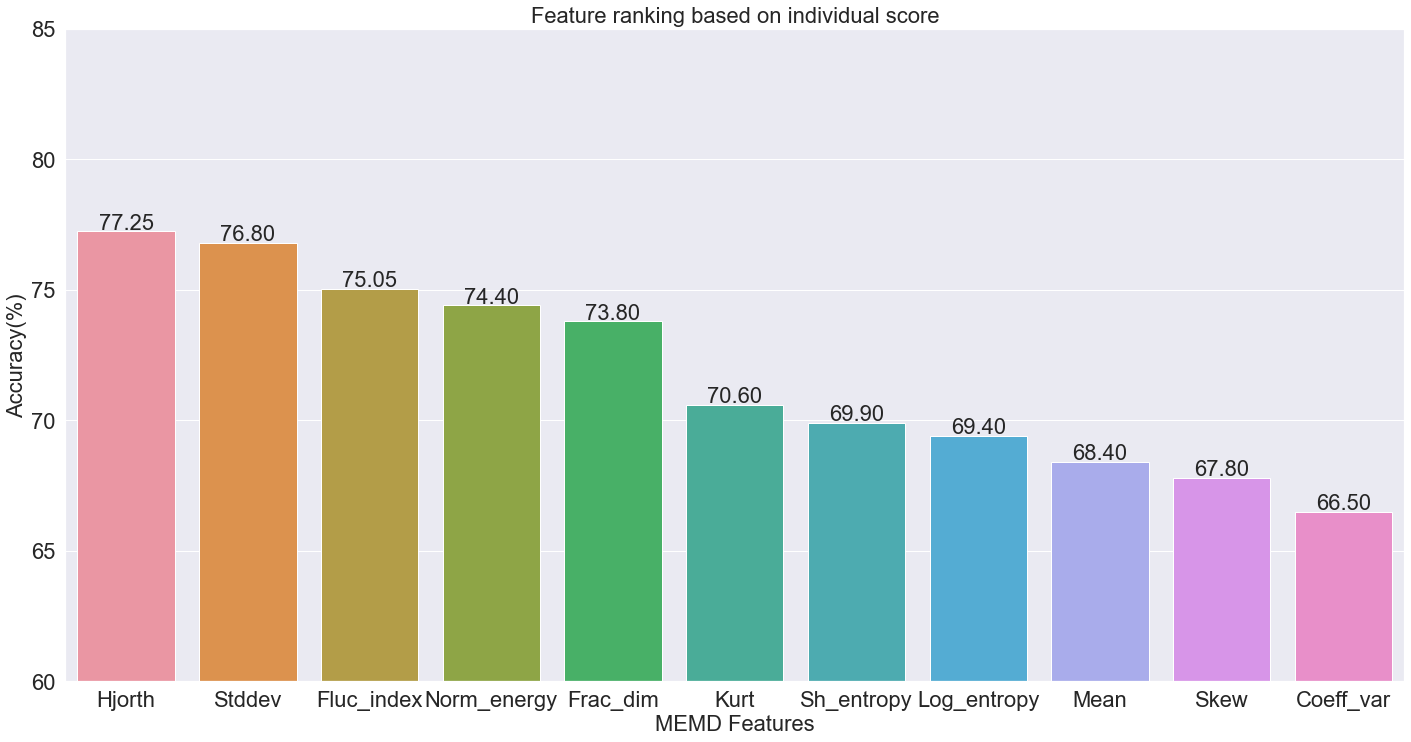}}
  \caption{Performance of individual MEMD features}
    \label{fig:res3}
\end{figure}

\begin{table}[h]
\caption{Mental State Detection Accuracy (\%) using Different Features}
\label{tab:table-name1}
\begin{center}
\begin{tabular}{|c|c|c|c|c| }
\hline
Feature & kNN & SVM  & XgBoost & RF-AdaBoost\\
\hline
DFT & 67.76 & 68.51 & 69.67 & 71.86\\
DWT & 73.26 & 70.14 & 74.44 & 75.23\\
MEMD & 90.54 & 95.14 & 96.12 & 97.08\\
MEMD-DWT & 92.83 & 95.17 & 96.92 & 97.74\\
\hline
\end{tabular}
\end{center}
\end{table}

As mentioned in Section III, $11$ features are derived from the selected IMFs. The performance accuracy of individual feature will indicate it's effectiveness on this task. In Fig. ~\ref{fig:res3}, accuracy obtained by each MEMD feature is illustrated where the features are ranked based on their individual performance. The combination of high-ranked features including Hjorth parameter, standard deviation, fluctuation index, normalized energy, fractal dimension, and kurtosis are the selected MEMD features which provides 97.08\% accuracy jointly. 

Table ~\ref{tab:table-name1} gives a comparison among the MEMD, DFT and DWT features with different classifiers. It reveals a significant advantage of the hybrid ensemble classifier over SVM, KNN and XgBoost. The hybrid classifier gives an accuracy of $97.08\%$ accuracy with the MEMD features, in comparison to $71.86\%$ and $75.23\%$ with DFT and DWT features respectively. This suggests that the high-oscillation IMFs carry more useful information than conventional tempo-spectral features. We suspect that the features extracted by linear wavelet decomposition are not suitable for EEG which have strong non-linear characteristics. The DWT features from subband 1 and 2 (covering 10-45 Hz) yield improved performance with $80.36\%$ accuracy. By combining those DWT and selected MEMD features, a slight improvement to $97.74\%$ has been achieved. To validate proposed methodology another publicly available dataset named DEAP was used that contains EEG signals of 32 healthy subjects where 40 video clips were used as stimuli. The participants provided a rating score in the range of 1 to 9 after watching each of 40 video clips. The two classes were labeled based on the score, i.e., “relax” ($<5$) and “working” ($> 5$). Our proposed approach achieved better performance with 84.56\% accuracy than the baseline approach in \cite {c16, c17}.

\subsection{Contributions of Localized Brain Lobes}

\begin{figure}[htb]
\begin{minipage}[b]{.48\linewidth}
  \centering
  \centerline{\includegraphics[width=4.2cm]{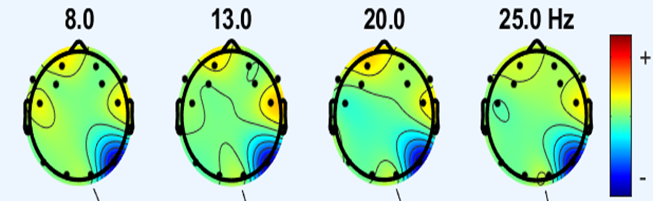}}
  \centerline{(a) Spectral-topoplot at relax}\medskip
\end{minipage}
\hfill
\begin{minipage}[b]{0.48\linewidth}
  \centering
  \centerline{\includegraphics[width=4.2cm]{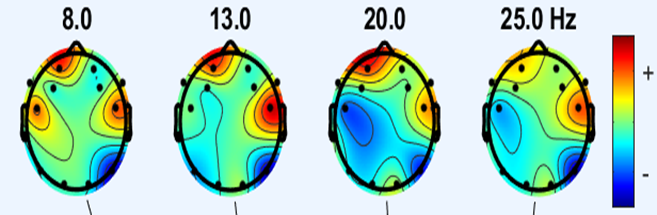}}
  \centerline{(b) Spectral-topoplot at working}\medskip
\end{minipage}
\caption{Brain region activation at specific mental state}
\label{fig:res1}
\end{figure}

\begin{figure}[htb]
\begin{minipage}[b]{1.0\linewidth}
  \centering
  \centerline{\includegraphics[width=6.8cm,height=3.6cm]{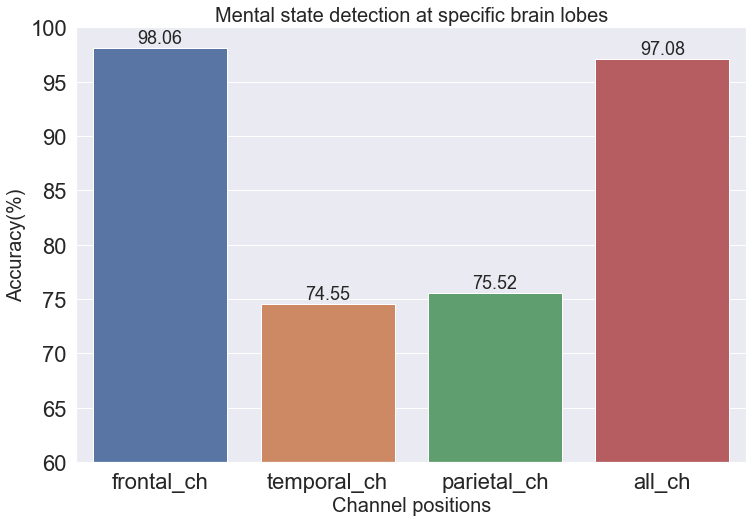}}

\end{minipage}
\caption{Mental state detection at specific brain regions}
\label{fig:res4}
\end{figure}

The spectral-topography visualizes the brain activation based on the power spectral density (PSD). Fig.~\ref{fig:res1} shows that the brain signal is more strongly triggered for working state than relax. The frontal brain lobe actively responds to mental state variation than the temporal and parietal lobes. To investigate the dominance of specific brain lobes, the region-specific MEMD features are analyzed. As illustrated in Fig.~\ref{fig:res4}, MEMD features extracted from EEG channels in the frontal brain region are most effective in mental state detection. This finding was only obtainable by functional magnetic resonance imaging (fMRI) in previous studies \cite {c18} and now our results used EEG to confirm that. 


\section{CONCLUSION}

This paper presents an approach to detecting mental states based on multivariate EMD. The experimental results show that MEMD features derived from lower-order IMFs are most effective in the task. The non-linear MEMD features demonstrate better performance than the commonly used DFT and DWT features. The detection performance attained by MEMD features from specific brain regions suggest that frontal EEG channels are more useful in detecting change of mental state than other channels.


\end{document}